\documentclass[12pt]{cernart}

\newcommand{\gevc }{GeV/$c$}
\newcommand{\DG }{$\left< {\Delta g}/{g} \right>_{x}$}
\newcommand{\Dg }{$\left< {\Delta g}/{g} \right>_{x}$~}
\newcommand{\dg }{\left< \frac{\Delta g}{g} \right>_{x}}
\newcommand{\AB} { $\left< A^{\gamma N}_{\rm B} \right> $}
\newcommand{\Ab} { $\left< A^{\gamma N}_{\rm B} \right> $~}
\newcommand{\GeV}{ \mathrm{GeV}}
\newcommand{\ua}{\uparrow }
\newcommand{\da}{\downarrow }

\usepackage{graphicx}
\usepackage{rotating}

\begin {document}
\dimen\footins=\textheight

\begin{titlepage}
\vspace{1cm}

\title{\LARGE
Gluon Polarisation in the Nucleon and Longitudinal Double Spin 
Asymmetries from Open Charm Muoproduction}
\vspace*{0.5cm}

\collaboration{COMPASS Collaboration}

\centerline{\today}

\vspace{2cm}
\begin{abstract}
The gluon polarisation in the nucleon has been determined by
detecting charm production via $D^0$ meson decay to charged $K$ and $\pi$ in polarised
muon scattering off a longitudinally polarised deuteron target. The data were taken by
the COMPASS Collaboration at CERN between 2002 and 2006 and corresponds
to
an integrated luminosity of 2.8~fb$^{-1}$. The dominant
underlying process of charm production is the photon--gluon fusion to a
$c \bar{c}$ pair. A leading order QCD approach gives an 
average gluon polarisation of \DG$=-0.49 \pm 0.27 (\mbox{stat}) \pm 0.11 (\mbox{syst})$ at
a scale $\mu^2\approx 13\,(\GeV/c)^2$ and at an
average gluon momentum fraction $\langle x_{\rm} \rangle
\approx 0.11^{}_{}$. The longitudinal cross-section
asymmetry for $D^0$ production is presented in bins of
the transverse momentum and the energy of the $D^0$ meson.
\end{abstract}

\vspace*{60pt}
PACS: 13.60.-r, 13.88.+e, 14.20.Dh, 14.70.Dj

keywords: Inelastic muon scattering; Spin; Asymmetry; Gluon
polarisation
\vfill
\submitted{submitted to Physics Letters}

\begin{Authlist}
{\large  The COMPASS Collaboration}\\[\baselineskip]
{\small
%
%
%
M.~Alekseev\Iref{turin_p},
V.Yu.~Alexakhin\Iref{dubna},
Yu.~Alexandrov\Iref{moscowlpi},
G.D.~Alexeev\Iref{dubna},
A.~Amoroso\Iref{turin_u},
A.~Austregisilio\IIref{cern}{munichtu},
B.~Bade{\l}ek\Iref{warsaw},
F.~Balestra\Iref{turin_u},
J.~Ball\Iref{saclay},
J.~Barth\Iref{bonnpi}
% ...
}
\end{Authlist}

%
%
\Instfoot{bielefeld}{Universit\"at Bielefeld, Fakult\"at f\"ur Physik, 33501 Bielefeld, Germany\Aref{f}}
\Instfoot{bochum}{Universit\"at Bochum, Institut f\"ur Experimentalphysik, 44780 Bochum, Germany\Aref{f}}
\Instfoot{bonniskp}{Universit\"at Bonn, Helmholtz-Institut f\"ur  Strahlen- und Kernphysik, 53115 Bonn, Germany\Aref{f}}
\Instfoot{bonnpi}{Universit\"at Bonn, Physikalisches Institut, 53115 Bonn, Germany\Aref{f}}
\Instfoot{brno}{Institute of Scientific Instruments, AS CR, 61264 Brno, Czech Republic\Aref{g}}
\Instfoot{burdwan}{Burdwan University, Burdwan 713104, India\Aref{h}}
\Instfoot{calcutta}{Matrivani Institute of Experimental Research \& Education, Calcutta-700 030, India\Aref{i}}
\Instfoot{dubna}{Joint Institute for Nuclear Research, 141980 Dubna, Moscow region, Russia}
\Instfoot{erlangen}{Universit\"at Erlangen--N\"urnberg, Physikalisches Institut, 91054 Erlangen, Germany\Aref{f}}
\Instfoot{freiburg}{Universit\"at Freiburg, Physikalisches Institut, 79104 Freiburg, Germany\Aref{f}}
\Instfoot{cern}{CERN, 1211 Geneva 23, Switzerland}
\Instfoot{liberec}{Technical University in Liberec, 46117 Liberec, Czech Republic\Aref{g}}
\Instfoot{lisbon}{LIP, 1000-149 Lisbon, Portugal\Aref{j}}
\Instfoot{mainz}{Universit\"at Mainz, Institut f\"ur Kernphysik, 55099 Mainz, Germany\Aref{f}}
\Instfoot{miyazaki}{University of Miyazaki, Miyazaki 889-2192, Japan\Aref{k}}
\Instfoot{moscowlpi}{Lebedev Physical Institute, 119991 Moscow, Russia}
\Instfoot{munichlmu}{Ludwig-Maximilians-Universit\"at M\"unchen, Department f\"ur Physik, 80799 Munich, Germany\AAref{f}{l}}
\Instfoot{munichtu}{Technische Universit\"at M\"unchen, Physik Department, 85748 Garching, Germany\AAref{f}{l}}
\Instfoot{nagoya}{Nagoya University, 464 Nagoya, Japan\Aref{k}}
\Instfoot{praguecu}{Charles University, Faculty of Mathematics and Physics, 18000 Prague, Czech Republic\Aref{g}}
\Instfoot{praguectu}{Czech Technical University in Prague, 16636 Prague, Czech Republic\Aref{g}}
\Instfoot{protvino}{State Research Center of the Russian Federation, Institute for High Energy Physics, 142281 Protvino, Russia}
\Instfoot{saclay}{CEA DAPNIA/SPhN Saclay, 91191 Gif-sur-Yvette, France}
\Instfoot{telaviv}{Tel Aviv University, School of Physics and Astronomy, 69978 Tel Aviv, Israel\Aref{m}}
\Instfoot{triest_i}{Trieste Section of INFN, 34127 Trieste, Italy}
\Instfoot{triest}{University of Trieste, Department of Physics and Trieste Section of INFN, 34127 Trieste, Italy}
\Instfoot{triestictp}{Abdus Salam ICTP and Trieste Section of INFN, 34127 Trieste, Italy}
\Instfoot{turin_u}{University of Turin, Department of Physics and Torino Section of INFN, 10125 Turin, Italy}
\Instfoot{turin_i}{Torino Section of INFN, 10125 Turin, Italy}
\Instfoot{turin_p}{University of Eastern Piedmont, 1500 Alessandria,  and Torino Section of INFN, 10125 Turin, Italy}
\Instfoot{warsaw}{So{\l}tan Institute for Nuclear Studies and University of Warsaw, 00-681 Warsaw, Poland\Aref{n} }
\Instfoot{warsawtu}{Warsaw University of Technology, Institute of Radioelectronics, 00-665 Warsaw, Poland\Aref{o} }
\Instfoot{yamagata}{Yamagata University, Yamagata, 992-8510 Japan\Aref{k} }
%
%
\Anotfoot{+}{Deceased}
\Anotfoot{a}{Also at IST, Universidade T\'ecnica de Lisboa, Lisbon, Portugal}
\Anotfoot{b}{On leave of absence from JINR Dubna}
\Anotfoot{c}{Also at Chubu University, Kasugai, Aichi, 487-8501 Japan$^{\rm j)}$}
\Anotfoot{d}{Also at KEK, 1-1 Oho, Tsukuba, Ibaraki, 305-0801 Japan}
\Anotfoot{e}{Also at GSI mbH, Planckstr.\ 1, D-64291 Darmstadt, Germany}
\Anotfoot{f}{Supported by the German Bundesministerium f\"ur Bildung und Forschung}
\Anotfoot{g}{Suppported by Czech Republic MEYS grants ME492 and LA242}
\Anotfoot{h}{Supported by DST-FIST II grants, Govt. of India}
\Anotfoot{i}{Supported by  the Shailabala Biswas Education Trust}
\Anotfoot{j}{Supported by the Portuguese FCT - Funda\c{c}\~ao para a Ci\^encia e Tecnologia grants POCTI/FNU/49501/2002 and POCTI/FNU/50192/2003}
\Anotfoot{k}{Supported by the MEXT and the JSPS under the Grants No.18002006, No.20540299 and No.18540281; Daiko Foundation and Yamada Foundation}
\Anotfoot{l}{Supported by the DFG cluster of excellence `Origin and Structure of the Universe' (www.universe-cluster.de)}
\Anotfoot{m}{Supported by the Israel Science Foundation, founded by the Israel Academy of Sciences and Humanities}
\Anotfoot{n}{Supported by Ministry of Science and Higher Education grant 41/N-CERN/2007/0}
\Anotfoot{o}{Supported by KBN grant nr 134/E-365/SPUB-M/CERN/P-03/DZ299/2000}
\end{titlepage}


\setcounter{footnote}{0}
\section{Introduction}

Pioneering experiments on the spin structure of the nucleon
performed in the seventies at SLAC \cite{vernon} were
followed by the EMC experiment at CERN which obtained
a surprisingly small quark contribution to the proton spin
\cite{emc}, in contrast to the naive expectation that the spin of
the nucleon is built mainly from valence quark spins \cite{leader}.
This result triggered
extensive studies of the spin structure of the nucleon 
in polarised lepton nucleon scattering experiments at CERN by
the SMC \cite{smc} and COMPASS \cite{compass}, at
SLAC \cite{e155_d}, at DESY \cite{hermes} and at JLAB
\cite{jlab} as well as in polarised proton--proton collisions at RHIC \cite{phenix,star}. 
As a result, 
the parton helicity distributions in the nucleon were extracted using
perturbative QCD analyses. The contribution of the quark spins
to the nucleon spin is now confirmed to be around 30\%, smaller than 
60\%, the value expected from the Ellis--Jaffe sum rule \cite{refa8}.
The reduction from the naive expectation of 100\% can be explained by
the relativistic nature of quarks (e.g.\ in the MIT bag model)
\cite{bass}.
However, due to the limited 
range in the four-momentum transfer squared, $Q^2$, covered by the experiments, the
QCD analyses
(e.g.\ \cite{compass}) show limited sensitivity to the gluon helicity
distribution as a function of the gluon momentum fraction $x$, $\Delta
g(x)$, and to its first moment, $\Delta G$. (The perturbative
 scale, $\mu^2$, in these QCD analyses is set to $Q^2$.)
The determination of  $\Delta g(x)$ from QCD evolution has therefore to be
complemented by direct measurements in dedicated experiments.  

The average gluon polarisation in a limited range of $x$, \DG,
has been determined in a model-dependent way from the
photon--gluon fusion (PGF) process
by HERMES \cite{hermes_highpt}, SMC \cite{SMC_highpt} and COMPASS
\cite{compass_highpt_lowq}. These analyses used events containing hadron pairs
with high transverse momenta, $p_{\rm T}$, (typically 1 to
  2~GeV/$c$) with respect to the virtual photon direction.
PYTHIA \cite{pythia} was used by HERMES and by COMPASS for the
  analysis
of small $Q^2$ events, while LEPTO \cite{lepto} was used in SMC and
the ongoing COMPASS analysis for $Q^2>1~(\GeV/c)^2$ events.
This method provides good statistical precision but relies
on Monte Carlo generators simulating QCD
processes. The measurements point towards a small
value of the gluon polarisation at $x\approx 0.1$. This is in line
with recent results from PHENIX \cite{phenix}
and STAR \cite{star} at RHIC.

Taking into account quark and gluon
orbital angular momenta, $L$, the nucleon spin projection
(in units of $\hbar$) is 
\begin{equation}
S_z=\frac{1}{2} = \frac{1}{2} \Delta \Sigma + \Delta G + L_z \,,
\label{spinsumrule}
\end{equation}
where $\Delta \Sigma$ is the first moment of the sum of the quark helicity
distributions. The decomposition of Eq.~(\ref{spinsumrule}),
however gauge dependent, is defined in the infinite momentum frame
where the quark parton model is valid.

Here we present a new result on \Dg from muon--deuteron
scattering.\footnote{The present result includes a larger data sample
  and an improved analysis method and thus supersedes the one given in
  Ref.~\cite{cernpp}.} 
The gluon polarisation is determined assuming that open-charm 
production is dominated by the
PGF mechanism yielding a $c\bar{c}$ pair which fragments
   mainly into $D$ mesons. This assumption is supported by the 
measurements of
   $F_2^c$ in the COMPASS kinematic domain \cite{emccharm} and further
discussed in \cite{reanal}.
This method has the advantage that in lowest order of the strong
  coupling constant there are no other
contributions to the cross-section; however, it is statistically
limited as will be shown in section~\ref{data-selection}.
In the present analysis only one
charmed meson is required in every event. This meson  is
selected through its decay in one of the two channels:
$D^{*}(2010)^+ \rightarrow D^0\pi^+_{\rm slow}\rightarrow K^-\pi^+\pi^+_{\rm
  slow}$  ($D^*$ sample) and $D^0\rightarrow K^-\pi^+$  ($D^0$
sample) and their charge conjugates.

\section{Experimental set-up}

The data
were collected between 2002 and 2006 with the COMPASS experiment at the
M2 muon beam line of the CERN SPS. A detailed description of the experiment
for the years 2002 to 2004 can be found in Ref.~\cite{spectrometer}.
For the 2006 data taking the polarised target
and the spectrometer were considerably upgraded.

The measurements were performed using a $\mu^+$ beam of 160~GeV/$c$. 
The beam muons originating from $\pi^+$ and $K^+$ decays are
naturally polarised with an average polarisation,
$P_{\mu}$, 
of about 80\% with a relative uncertainty of 5\% \cite{beampol}.
The momentum of each incoming muon is measured upstream of the experimental 
area  with a precision of $\Delta p/p\le 1\%$ in a 
beam momentum station consisting of layers of scintillators. The incoming muon  direction and position is measured
with a detector telescope in front of the target. A precision of 30~$\mu$rad
is obtained for the track direction.

The polarised $^6$LiD target is housed in a superconducting
solenoid with a polar angle aperture of 70~mrad
in 2002 to 2004. The target
consisted of two 60~cm long cells (upstream $u$,
downstream $d$), separated by 10~cm, longitudinally polarised with 
opposite orientations. 
The spin directions were reversed every eight hours by rotating the
field of the target magnet system. 
The target was upgraded in 2006 with a new 
solenoid with an aperture of 180~mrad.
To reduce the systematic errors due to the 
different spectrometer acceptances for the upstream and downstream
cells, a 3-cell target configuration was installed.
A central 60~cm long cell is placed in-between two 30~cm long cells
with polarisations opposite to the central one.\footnote{In 2006 $u$
  and $d$ stand for the central target cell and for the sum of the outer
  target cells,
  respectively.} In this set-up the average acceptances 
for both spin directions are very similar and therefore
the magnetic field was rotated only once per day.
The average target polarisations, $P_{\rm t}$, were $50$\% with a
relative uncertainty of 5\%.
The dilution factor $f$, accounting for the fraction of polarisable
nucleons in the target, 
is about 0.4, since the $^6$Li nucleus basically consists of a $^4$He core
plus a deuteron. 
The exact value of $f$ is kinematics dependent and is calculated as described
in Ref.~\cite{compass_A1}. Its relative uncertainty is 5\%.

The two-stage COMPASS spectrometer is designed to reconstruct the scattered
muons and the produced hadrons in a wide momentum range. Particle
tracking is performed using several stations
of scintillating fibres, micromesh gaseous chambers and gas electron
multiplier chambers for the small angles tracks. Large area tracking devices
comprise gaseous detectors (drift chambers, straw tubes and multiwire
proportional chambers). The detectors are placed around the two spectrometer
magnets. The direction of the tracks reconstructed at an interaction point in
the target is determined with a precision better than 0.2~mrad and the momentum
resolution for charged tracks detected in the first spectrometer is about 1.2\% whereas
is it about 0.5\% in the second spectrometer. The achieved
longitudinal vertex
resolution varying from 5~mm to 25~mm along the target allows
assigning each event to a particular target
cell, i.e. a specific target spin direction. For 2006 the tracking
systems in the first stage were adapted to match the increased aperture of the
polarised target magnet.
The trigger is formed by
several hodoscope systems supplemented by two hadron calorimeters.
Muons are identified downstream of the hadron absorbers. 
A Ring Imaging CHerenkov counter (RICH) with a C$_4$F$_{10}$ radiator 
is used in the first spectrometer stage for charged particle
identification. 
It is equipped with multiwire proportional chambers with CsI photocathodes 
to detect the UV Cherenkov photons. 
The RICH, too, underwent a considerable upgrade 
for the 2006 data taking. In the central part, the
photon detectors were replaced by multi-anode photomultiplier tubes,
yielding considerably higher photon detection efficiency 
along with a much faster response. For the outer parts the
readout electronics was refurbished, allowing a significant reduction
of the background. The data taking amounted to 40 weeks in 2002 to
2006 and corresponds to an integrated luminosity of 2.8~fb$^{-1}$.
\section{Data selection}
\label{data-selection}

In the present analysis the selection procedure required  an incoming
muon, a scattered muon, an interaction vertex
in the target and at least two additional tracks.
The kinematic variables like the four-momentum transfer squared $Q^2$, the
relative energy transfer $y$, and the Bjorken variable $x_{\rm
  Bj}=Q^2/2MEy$,
where 
$M$ is the nucleon mass and E the incident muon energy, are calculated from the
four-momenta of the incident and scattered muon.
No kinematic cuts are applied on $Q^2$, $y$ or $x_{\rm Bj}$. Thus the selected
data sample includes the events with an interaction vertex from quasi-real
photo-production $Q^2\approx m_{\mu}^2 y^2/(1-y)$ to a $Q^2$ of about 100~$(\GeV/c)^2$. Note that all the
events are in the deep inelastic region, i.e. the invariant mass
of the final state, $W$, is larger than 4~GeV/$c^2$.

The $D^0$ mesons are reconstructed through their $K\pi$ decay
which has a branching ratio of 3.9\%. Due to
multiple Coulomb scattering of the charged particles in the 
solid state target the spatial resolution of the vertex reconstruction
is not sufficient to separate the $D^0$ production and
decay vertices.
The $D^0$ mesons are selected using the
invariant mass of their
decay products. 

To reduce the large combinatorial background only identified
$K\pi$ pairs are used.
The identification in the RICH starts from reconstructed tracks with
measured momenta. The likelihood for different mass hypotheses and for a
background hypothesis 
are computed for each track, using the angles between the track and the
detected Cherenkov photons. The likelihood functions, used in this
computation, were defined from the corresponding expected angular
distribution of photons; the expected distribution for background was
obtained using a sample of photons not associated to reconstructed tracks.
Particles are identified as kaons or pions on the basis of the
likelihood associated to the pion, kaon,
proton and background hypotheses.
The procedure restricts the studied events to a sample with at least 
one kaon and one
pion of momenta exceeding the Cherenkov threshold of 
$9.1\,\GeV/c$ and  $2.5\,\GeV/c$,
respectively. Simulations using the AROMA \cite{aroma} generator and a
full spectrometer simulation based on GEANT 
have shown that about 70
\% (90\%) of kaons (pions) coming from $D^0$ decays exceed this
threshold
for the reconstructed sample.

\begin{figure}[tb]
\begin{center}
\includegraphics[width=0.8\hsize,clip]{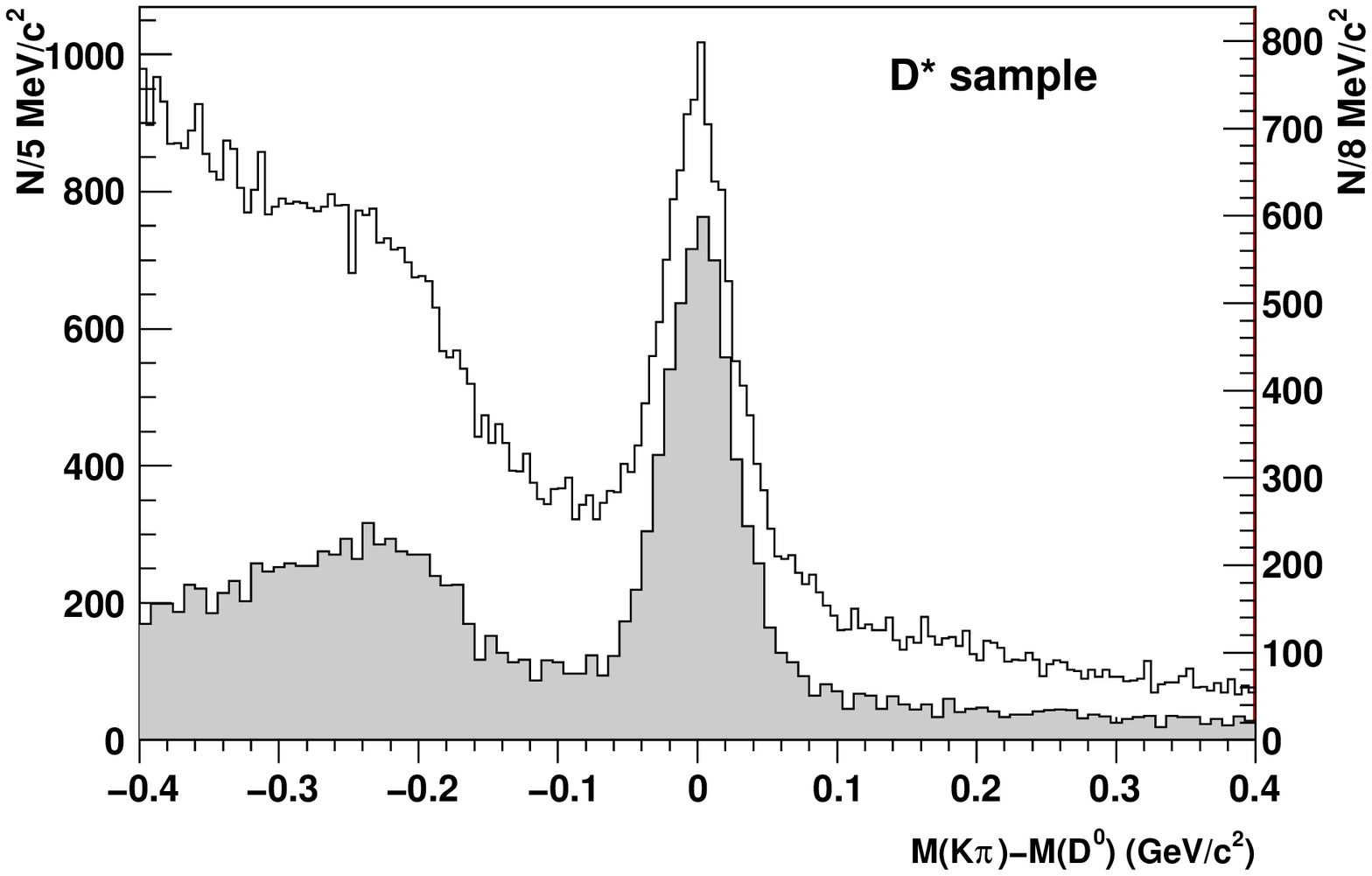}
\includegraphics[width=0.8\hsize,clip]{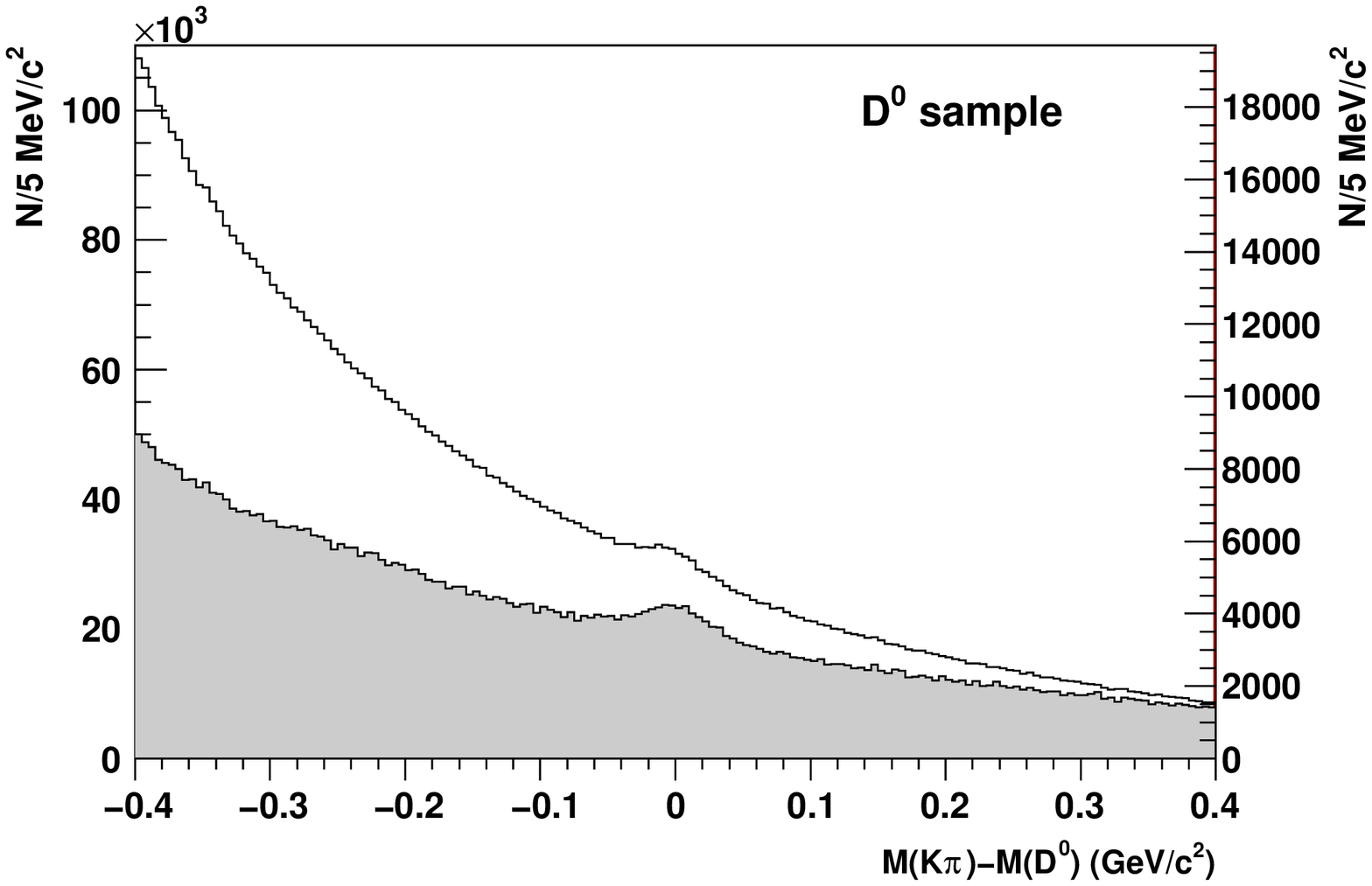}
\end{center}
\caption{\small Invariant mass distributions of the $K\pi$ pairs for
  the $D^*$ sample (upper plot) and the $D^0$ sample (lower plot). 
The non-shaded histograms (left scale) show the total event samples
while the shaded ones (right scale) show the events in the 
highest bin of $(S/B)_{\rm par}$.
}
\label{d0fit_all}
\end{figure}

All events have to satisfy a kinematic cut: $z>0.2$,
 where $z$ is the fraction of the energy
of the virtual photon carried by the $D^0$ meson candidate.
They are further divided into a $D^*$ and a $D^0$ sample, analysed
independently. 
In the former one an additional track with a proper charge, a slow pion
candidate, is demanded at the vertex. RICH likelihoods, used to reject 
electrons from those candidates, reduce the combinatorial background 
by a factor two. Furthermore, in the case of the $D^*$, a cut on the
mass difference is imposed, $3.2 \mathrm{\ MeV/}c^2 < M_{K\pi\pi_{slow}} -
M_{K\pi} - M_{\pi}< 8.9 \mathrm{\ MeV/}c^2$, where $M_{K\pi\pi_{slow}}$
and $M_{K\pi}$ are the masses of the $D^*$ and the $D^{0}$
candidates, respectively.
Finally it was demanded that 
$|\mathrm{cos}\theta^*|<0.9$ for the $D^*$ sample and 
$|\mathrm{cos}\theta^*|<0.65$ for the $D^0$,
where $\theta^*$ is the decay angle in the $D^{0}$ c.m. system
relative to the $D^0$ flight  direction. 
  The events entering the $D^*$ sample are not used in the $D^0$ sample.
The resulting mass spectra for the $D^0$ and $D^*$ samples with one
$K \pi$ pair in the mass range $-400$~MeV/$c^2
< M_{K\pi} - M_{D^0}  < 400$~MeV/$c^2$ are
displayed 
in Fig.~\ref{d0fit_all}. A signal to background ratio in the signal region
of about 1 is obtained for the $D^*$ sample and of about 0.1 for the $D^0$
sample with a mass resolution of about 22~MeV/$c^2$ and 25~MeV/$c^2$, 
respectively.
The number of $D^0$ mesons is about 8,700 and 37,400 
in the $D^*$ and the $D^0$ samples.

For the final event samples the mean value of $Q^2$ is
  $0.65~(\GeV/c)^2$, $x_{\rm Bj}$ ranges from
$1\cdot 10^{-5}$ to 0.6 with a mean value of 0.04 and $y$ from 0.1 to 1 
with a mean value of 0.55.
Note that the perturbative scale for the selected events is not
  given by $Q^2$, but by the transverse mass of the charmed quarks,
$M^2_{\rm T} = 4(m^2_c + p^2_{\rm T})$.

\section{Method}

This section describes the determination of the gluon polarisation 
from the event samples collected in two different spin configurations
and target cells. The same method is used in section \ref{asym_det} for the
asymmetry determination.
The number of events collected  in a given target cell and
time interval is
\begin{equation}
\label{N}
 \frac{{\rm d}^k N}{{\rm d}m\, {\rm d}X}= a \phi n (s+b)  
  \left [1 + P_{\rm t} P_{\mu} f \left(
          \frac{s}{s+b}  A^{\mu N\rightarrow \mu' D^0 \rm{X}}+ 
         \frac{b}{s+b}
  A_{\rm B} \right)\right ]\,\,.
\end{equation}
Here, $A^{\mu N\rightarrow \mu' D^0 {\rm X}} = 
(\sigma^{\ua \da} - \sigma^{\ua \ua})/(\sigma^{\ua \da} + \sigma^{\ua
    \ua})$, where the arrows
indicate the relative beam and target spin orientations, 
 is the longitudinal double spin cross-section
 asymmetry
of the events in the central peak of Fig.~\ref{d0fit_all} and $A_{\rm
  B}$  is the corresponding 
asymmetry originating from the combinatorial background events in the
mass spectra. Also,
$m\equiv M_{K\pi}$, and $X$ denote a set of kinematic variables
describing an event ($Q^2$, $y$, $z$...), while
$a$, $\phi$ and $n$ are the spectrometer acceptance, the
integrated incident muon flux and the number of target nucleons,
respectively.
The differential unpolarised cross-sections of signal and
background events folded with the experimental resolution as a function of
$m$ and $X$ are represented by $s=s(m,X)$ 
and $b=b(m,X)$, respectively. 
The ratio ${s}/(s+b)$ 
will be called ``signal purity''.
In the present analysis the background is a combinatorial
  background and the signal purity can be extracted from the data
using the invariant mass distributions of Fig.~\ref{d0fit_all}. This is
in contrast to the high-$p_{\rm T}$  analyses, where the physical
background has to be estimated using a
Monte Carlo simulation  (MC)~\cite{hermes_highpt, SMC_highpt,compass_highpt_lowq}. 
Information on the gluon polarisation is contained in 
$A^{\mu N\rightarrow \mu' D^0 {\rm X}}$
which can be decomposed in LO QCD as
\begin{equation} \label{asym-factor}
 A^{\mu N\rightarrow \mu' D^0 {\rm X}}(X)=a_{\rm LL}(X)\, \frac{\Delta g}{g}(X)\,. 
\end{equation}
Here $a_{\rm LL}$ is the analysing power of the 
$\vec \mu \vec g \rightarrow \mu' c \bar c$ process which includes
the so-called depolarisation factor $D$ accounting for the
polarisation
transfer from the lepton to the virtual photon.
The background asymmetry $A_{\rm B}$
can be written as the product
of the virtual photon asymmetry and the depolarisation factor
$A_{\rm B} = D A^{\gamma N}_{\rm B}$ and is assumed to be independent
of $m$.

In the present analysis the average gluon polarisation \Dg and the
average background asymmetry \Ab are determined simultaneously
as weighted averages over the accessible kinematic range.
This method does not require an arbitrary selection of mass windows
for the signal and background regions as in the
classical side-band subtraction method.
Moreover, it yields a smaller statistical error compared to the latter,
reaching practically the lower bound
of the unbinned likelihood method \cite{pretz}.  
This is achieved by weighting every event with its 
analysing power $a_{\rm LL}(X)$.
The same procedure is applied for $A^{\gamma N}_B$.
The weighting factors are thus
\begin{equation}
\label{weight}
w_{\rm S} = P_{\mu} f \frac{s}{s+b} a_{\rm LL} \,\,,\,\,
    w_{\rm B} = P_{\mu} f \frac{b}{s+b} D\,\,.
\end{equation}
The target polarisation $P_t$, as a time dependent quantity, is not included into 
the weights because including it may generate false asymmetries.
Note that all events in the mass window $-400$~MeV/$c^2< M_{K\pi}  - M_{D^0} <
400$~MeV/$c^2$ of Fig.~\ref{d0fit_all}
are used. Since the factor $s/(s+b)$ in $w_S$ vanishes
for events far away from the central peak, these events do not contribute
significantly to \DG, but contribute to the determination of \AB.

By considering sums over the different event samples eight equations are
derived from Eq.~(\ref{N})
\cite{pretz_hab}
\begin{equation}\label{W}
\sum_{i=1}^{N_t} w_{{\rm C},i}=\alpha^t_{\rm C}\left(1+
\beta^t_{\rm C} \dg +\gamma^t_{\rm C} \left< A_{\rm B}^{\gamma
  N}\right > \right ) 
\end{equation}
 \begin{eqnarray}
\beta^t_{\rm C}
\approx \frac{\sum_i^{N_t} P_{{\rm t},i} w_{{\rm S},i} w_{{\rm 
C},i}}{\sum_i^{N_t} w_{{\rm C},i}}\,\,,\,\,
\gamma^t_{\rm C}
\approx \frac{\sum_i^{N_t} P_{{\rm t},i} w_{{\rm B},i} w_{{\rm 
C},i}}{\sum_i^{N_t}
w_{{\rm C},i}} 
\end{eqnarray}
for the two target cells before ($t=u,d$) and 
after ($t=u',d'$) the target spin reversal, 
once weighted with $w_{\rm S}$ and once with $w_{\rm B}$ (${\rm C=S,B}$).  
Here $N_t$ is the number of events observed in cell $t$.
These eight equations contain 10 unknowns which are
\DG, \Ab and eight acceptance factors $\alpha_{\rm C}^{t}=\int a^t \phi^t
n^t (s+b)  w_{\rm C}\, {\rm d}X$.

Assuming that possible acceptance variations affect the upstream and
downstream cells in the same way, i.e.
   ${\alpha_{\rm C}^{u}/ \alpha_{\rm C}^{d}}
    =    {\alpha_{\rm C}^{u'}/ \alpha_{\rm C}^{d'}}$,
reduces the number of unknowns to eight.
With an extra, much weaker assumption that signal and background 
events from the same target cell
are affected in the same way by the acceptance variations,
one arrives at a system of eight equations with seven unknowns.
Possible deviations from the above assumptions may generate false
asymmetries which are included in the systematic 
error. Using the set of eight equations (see Eq.~(\ref{W})), 
the gluon polarisation \Dg and the background asymmetry 
\Ab are determined with
a standard least square minimisation procedure taking into account
the statistical correlation between the number of events in a given
target cell weighted by $w_{\rm S}$ and
by $w_{\rm B}$. 
The analysis is performed independently for the $D^*$ and
$D^0$ samples. 

\begin{figure}[tbp]
\begin{center}
\includegraphics[width=0.8\hsize,clip]{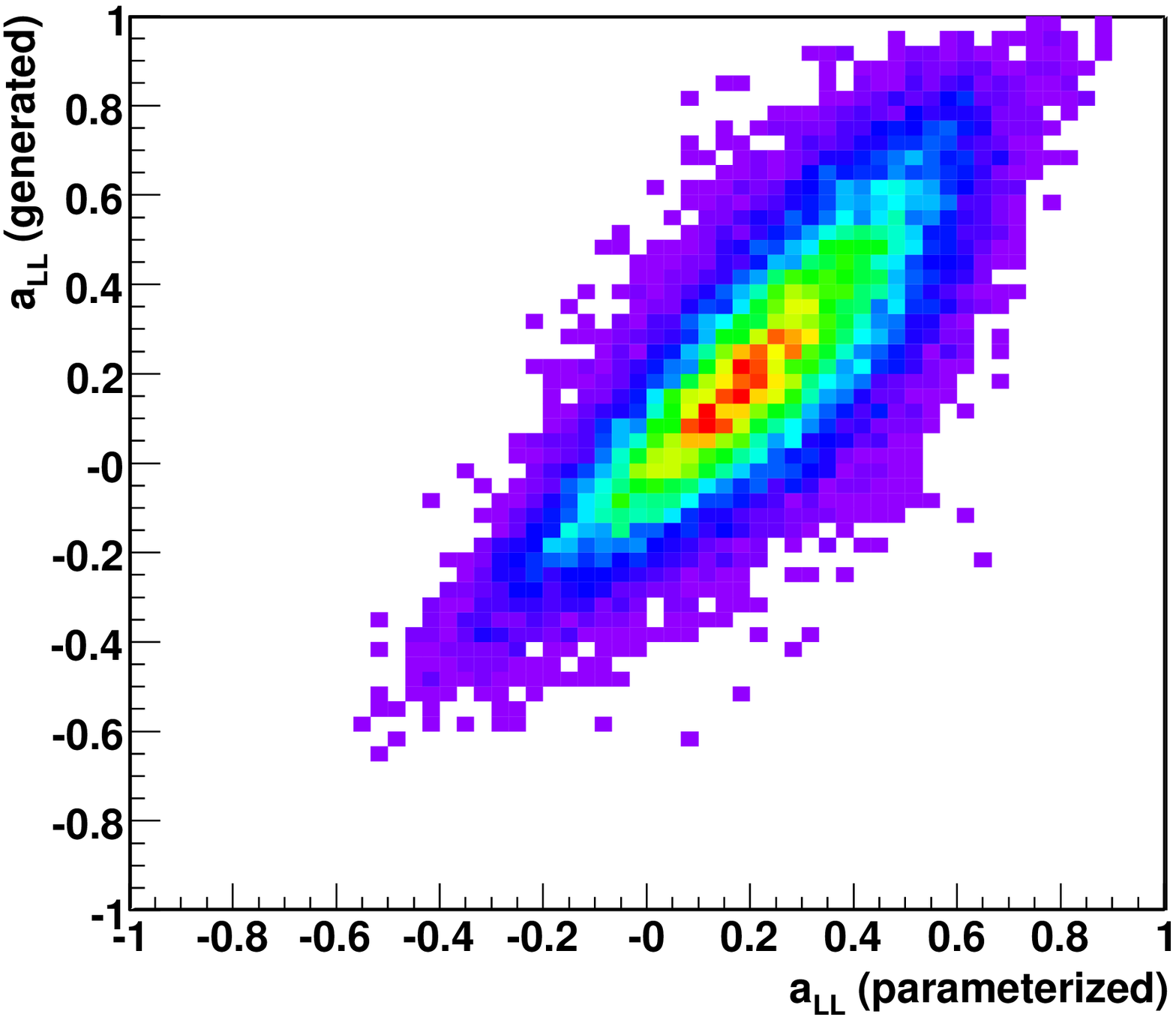}
\end{center}
\caption{
Correlation between the generated analysing power $a_{\rm LL}^{gen}$ and the analysing power parameterised by neural
network $a_{\rm LL}^{par}$. 
}
\label{a_ll}
\end{figure}

The quantities $P_{\rm t}$,
$P_{\mu}$,
$a_{\rm LL}$ and $S/(S+B)$ are obtained as follows.
For $P_{\rm t}$, values averaged over about one hour 
of data taking are used, a timescale over which the assumption
of a stable target polarisation was shown to be justified.
The beam polarisation $P_{\mu}$ is parameterised as a function of the momentum
which is measured for each incoming muon.
The photon--gluon analysing power, $a_{\rm LL}(X)/D$, 
is parameterised in terms of measured
kinematic variables. It depends on partonic variables not accessible
experimentally and is obtained using  a neural
network \cite{neuralnet} trained on a Monte Carlo sample for $D^*$ mesons.
For this purpose PGF events were generated
with AROMA \cite{aroma} in leading order QCD,  
processed by GEANT to simulate the response of
the detector and finally reconstructed like real events.
It was checked that the MC simulation describes the background
subtracted data distributions in $z$ and $p_{\rm T}$ sufficiently well.
The scale, $\mu$, used in the MC was chosen as the  
transverse mass of the produced charmed quark pair, and is sufficiently large
to justify the perturbative approach.
The correlation between the generated $a_{\rm LL}$
and  the parameterised $a_{\rm LL}$ is 81\% (see Fig.~\ref{a_ll}).
The same parameterisation is valid for the $D^0$
and the $D^*$ samples.

Finally, the signal purity, $s/(s+b)$, as a
function of the invariant mass for each
event, is determined from a fit of the invariant mass distributions of
the $D^*$ and $D^0$ samples. In this fit the signal is described by 
a Gaussian distribution.
In the $D^*$ case the background function is the sum of an exponential and a
Gaussian, the latter added to describe the reflection of the
$D^{\,0}\,\rightarrow\, K\,\pi\,\pi^{\,0}$ decay, where the $\pi^{\,0}$
meson is not observed. In the $D^0$ case the background is described by
the sum of two exponential distributions. Note that not only the variation of the
signal purity (or $s/b$) with the mass, but also with other characteristics of
the event, is taken into account. 
This is achieved by a method \cite{phd-robinet} based on a multivariate approach starting
with a parameterisation of the signal-to-background
ratio, integrated over a window around the $D^0$ mass, $(S/B)_{\rm par}$.
The window is of $\pm 40$ MeV/$c^2$ for the $D^*$ sample and $\pm 30$
MeV/$c^2$ for the $D^0$ sample. 
The parametrisation is the product of 10 functions, each one depending on
one of the 10 variables describing the event kinematics and the RICH response.
Typically six bins are defined in each of
the variables and the mass spectra are fitted in each bin of each variable
to provide the values of the $S/B$ ratios using the signal and background
functions described above.
Each of the 10 variables is considered successively and the parameters
of the corresponding function are adjusted to reproduce the
$S/B$ ratios in all bins in this variable. Adjusting the
parameters for one variable affects the agreement obtained for previous
variables and thus the adjustment procedure has to be repeated until
convergence is reached and all $S/B$ ratios are reproduced simultaneously.

Using this parametrisation, each sample ($D^*$ and $D^0$) is split into 
intervals of
$(S/B)_{\rm par}$ and the mass spectrum is fitted separately in each of them. 
As an illustration the invariant mass spectra obtained in the highest 
interval of $(S/B)_{\rm par}$ are compared in Fig.~\ref{d0fit_all} to 
those obtained for the full samples.
The signal purity for each event is obtained from the fit to the mass 
spectrum
in the interval of $(S/B)_{\rm par}$ containing the event and this value is
adjusted to the exact value of $(S/B)_{\rm par}$ for this event.
To validate the procedure the fit in each $(S/B)_{\rm par}$ interval 
is integrated
over the window around the mass peak to obtain the $S/B$ value and compared 
with the average value obtained
from the parametrisation. The consistency obtained guarantees that
using the $(S/B)_{\rm par}$
in the event weights does not introduce a bias. 
In addition, it is checked that weighting the wrong-charge background 
($K^-\pi^+\pi^-_{\rm slow}$ and charge conjugates) with
the parametrised values of the mass averaged signal purity,
$[S/(S+B)]_{\rm par}$ ,
does not generate any artificial peak at $M_{K\pi}=M_{D^0}$.


\section{Results for the gluon polarisation}

A value for \Dg is obtained for each of the 40 weeks of data taking
separately for the $D^0$ and the $D^*$ sample. 
The results \DG=$-0.421\pm 0.424 (\mbox{stat})$ for the $D^0$ and
\DG$-0.541\pm 0.343 (\mbox{stat})$ for the $D^*$ sample, are the weighted mean of these values.
The resulting background asymmetries, $\left< A^{\gamma N}_{\rm B}
\right> = 0.003 \pm 0.004$ for the $D^0$ sample and $\left< A^{\gamma N}_{\rm B}
\right> = 0.062 \pm 0.042$ for the $D^*$ sample,
are consistent with zero. Assuming that ${\Delta g}/{g(x)}$
is approximately linearly dependent on $x$ in the range covered, 
\Dg gives a measurement of $\Delta g/g(\langle x \rangle)$, where
$\langle x \rangle$ is calculated using the
signal weights. This assumption is supported by the results of
the COMPASS QCD analysis~\cite{compass}.

\begin{table}[tp]
\caption{\small Systematic error contributions to \Dg for $D^0(D^*)$ channels.}
\label{tab:D0}
\begin{center}
\begin{tabular}{lc||lc}
\hline
\hline
source & $\delta ( \langle \frac{\Delta g}{g} \rangle_x )$ & source &$\delta 
(\frac{\Delta g}{g})$ \\
\hline
\hline
False asymmetry   & $ 0.05(0.05)$&Beam polarisation $P_{\mu}$   &  0.02  \\
$S/(S+B)$   & $0.07(0.01)$&Target polarisation $P_{\rm t}$   & 0.02\\
$a_{\rm LL}$   & $0.05(0.03)$&Dilution factor $f$ & 0.02 \\
\hline
\multicolumn{4}{c}{Total error~~~~0.11(0.07)}\\
\hline
\hline
\end{tabular}
\end{center}
\end{table}

\begin{figure}[tbp]
\begin{center}
\includegraphics[width=0.8\hsize,clip]{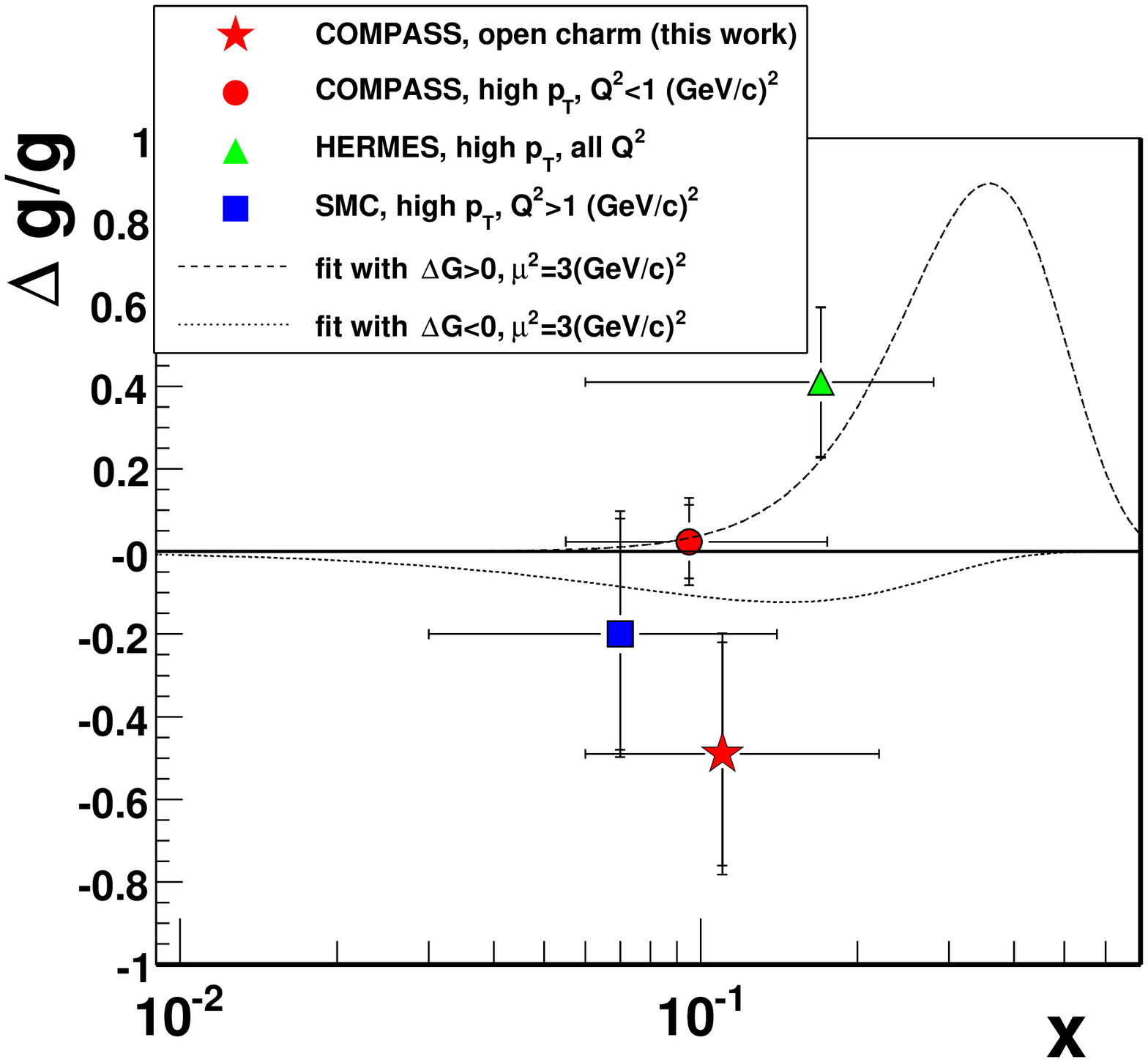}
\end{center}
\caption{\small Compilation of the \Dg measurements from
open charm and high-$p_{\rm T}$ hadron pair production by COMPASS 
\cite{compass_highpt_lowq},
SMC \cite{SMC_highpt} and HERMES \cite{hermes_highpt} as a function of $x$.
The horizontal bars mark the range in $x$ for each measurement, the vertical
ones give the statistical precision and the total errors (if available). 
The open charm measurement is at
a scale of about $13~(\GeV/c)^2$, other measurements at 3~$(\GeV/c)^2$.
The curves display two  parameterisations from the COMPASS 
QCD analysis at NLO \cite{compass}, with $\Delta G > 0$ (broken line)
and with $\Delta G < 0$ (dotted line).}
\label{DGG_all}
\end{figure}

The major contributions to the systematic uncertainty are listed in 
Table~\ref{tab:D0}. The contributions from $P_{\mu}$, $P_{\rm t}$ and $f$ 
are discussed with more detail in Ref.~\cite{compass}. 
To study the influence of false asymmetries, the final samples from
Fig.~\ref{d0fit_all} were subdivided 
into two samples using criteria related to the experimental apparatus, 
e.g.\ kaons going to the upper or to the lower spectrometer parts. 
The resulting asymmetries were found to be compatible
within their statistical accuracy, thus no false asymmetries were observed.
An upper limit of the contribution of time dependent acceptance effects
to the systematic uncertainty was 
estimated from the dispersion
of the values for \Dg and \Ab for the 40 weeks of data taking.
Assuming that possible detector 
instabilities are similar for background and signal events and applying the
method used in Ref.~\cite{compass} leads to a conservative limit of $0.05$ for both
decay channels.

Varying the procedure to build the parameterisation of $s/(s+b)$, 
and in particular the functional form of the background fit, results in 
an error on \Dg of 0.07 and 0.01 
for the $D^0$ and the $D^*$ sample, respectively.
As expected, the uncertainty on 
$s/(s+b)$
is larger for the $D^0$ case, where the signal-to-background ratio  is smaller.
To estimate the influence of the simulation parameters,
i.e.\ charmed quark mass (varied from 1.3~GeV/$c^2$ to 1.6~GeV/$c^2$),
parton distribution functions and scales (varied by a factor of 8), 
MC samples with different parameter
sets were generated and $a_{\rm LL}$ was recalculated,
resulting in an uncertainty on \Dg of 0.05 and 0.03 for the $D^0$ 
 and the $D^*$ sample, respectively.
Other contributions, like radiative
corrections and event migration between target cells,
were studied and found to be negligible. 

The final value is the weighted mean of the two values for the $D^*$
and the $D^0$ sample and amounts to 
\begin{equation}\label{finalres}
\dg =-0.49\pm 0.27(\mbox{stat})\pm 0.11(\mbox{syst})
\end{equation}
in the range of $0.06<x< 0.22$ with  $\langle x \rangle \approx 0.11$, 
and a scale $\langle
\mu^2\rangle \approx 13~(\GeV/c)^2$. The contributions to the
systematic uncertainty for each sample are
added in quadrature
to obtain the total error, 0.11 and 0.07 for the $D^0$ and $D^*$
sample, respectively.
The larger value is chosen as a conservative estimate 
of the final error in Eq.~(\ref{finalres}).
 
In Fig.~\ref{DGG_all} the above result is compared to other measurements of
\Dg and to two parametrisations from the NLO QCD analysis
of the world data on the polarised structure function $g_1(x,Q^2)$, performed
by COMPASS \cite{compass}: with $\Delta G > $0 (broken line) and with $\Delta G < 0$
(dotted line). The present result is consistent with previous measurements
favouring small values of \DG. Note that $Q^2$ is the scale for the
analysis of the SMC \cite{SMC_highpt} measurement and the QCD analysis
\cite{compass}. The scale of 
the present result is given by the transverse mass of the charmed quarks
$\mu^2=M^2_{\rm T} \approx 13~(\GeV/c)^2$.
The other experimental points in Fig.~\ref{DGG_all} are given at
$\mu^2\approx 3~(\GeV/c)^2$.

\section{Asymmetry determination}
\label{asym_det}

The data described in sections 2 and 3 also allow for the
  determination of the
virtual photon asymmetry for $D^0$ production, 
$A^{\gamma N \rightarrow D^0 \rm{X}} = A^{\mu N\rightarrow \mu'
  D^0}/D$.
In contrast to \Dg this asymmetry is
independent of the interpretation in LO QCD.
The asymmetry averaged over the full kinematic range would
be largely diluted because of the large dispersion of 
$a_{LL}$.
The asymmetry $A^{\gamma N \rightarrow D^0 \rm{X}}$ is thus extracted
in bins of the transverse momentum of the $D^0$ with
respect to the virtual photon, $p_{\rm T}^{D^0}$, and the energy of the $D^0$ in
the laboratory system, $E_{D^0}$. 
The bins were chosen such that the variation of 
$a_{\rm LL}/D$ within each bin
is small compared to the variation over the whole sample.
In principle $A^{\gamma N\rightarrow D^0 \rm{X} }$ also depends on the inclusive
variables $y$ and $Q^2$, but an additional binning is not necessary because the
dependence is very weak.
This is clearly
seen in LO, where $A^{\gamma N\rightarrow D^0 \rm{X} } =(a_{\rm LL}/D) \, \Delta g/g$.
In a given bin in $p_{\rm T}^{D^0}$ and $E_{D^0}$ the factor $(a_{\rm LL}/D)$
is almost independent of $y$ and $Q^2$, and the same is true for $\Delta g/g$.

The asymmetry $A^{\gamma N\rightarrow D^0 \rm{X} }$ is  obtained in exactly the
same way as \DG, except that the factor $a_{\rm LL}$ is replaced by $D$ in the
definition of the signal weight in Eq.~(\ref{weight}), i.e.~$w_S=P_\mu f D {s}/({s+b})$.
This provides $A^{\gamma N\rightarrow D^0 \rm{X}}(\langle p_{\rm T}^{D^0}\rangle, \langle
E_{D^0}\rangle)$ under the assumption that the bins of $p_{\rm T}^{D^0}$ and
$E_{D^0}$ are small enough.
It was verified that this approximation and the independence on $y$
and $Q^2$ are
well fulfilled for the cross-section evaluated in LO QCD. 
At higher orders, the variation of the cross-section 
are expected to be similar and thus the approximations to remain
valid.

Table~\ref{asym_results_d0ds} gives
$A^{\gamma N\rightarrow D^0 \rm{X}}$ averaged over the $D^0$ and $D^*$ sample
in each $(p_{\rm T}^{D^0},E_{D^0})$
bin, together with the average of several kinematic variables. 
All averages are calculated with the weight $w_S=P_\mu f D {s}/({s+b})$.
The muon-nucleon asymmetry $A^{\mu N\rightarrow
  \mu' D^0 \rm{X}  }$ can be obtained by multiplying 
$A^{\gamma N \rightarrow D^0 \rm{X}}$ by $D(\langle X\rangle)$.
Both asymmetries can be used in global NLO QCD fits to constrain the
values of $\Delta g(x)$.

As a cross-check we have calculated \Dg from $A^{\gamma N\rightarrow D^0 \rm{X}}$
in each bin by dividing the asymmetry by the corresponding $a_{\rm LL}/D$. 
Combining all bins we got a result consistent
with the result in Eq.~(\ref{finalres}), with an increase of 5\% in the
statistical error.
The contributions to the systematic error listed in Table~\ref{asym_results_d0ds} contribute
as well to the systematic error of the asymmetries, except for the
contribution of $a_{LL}$. This leads to a relative systematic uncertainty of 20\%
for $A^{\gamma N\rightarrow D^0 \rm{X}}$ which is 100\% correlated
between the bins.

\begin{sidewaystable}
\begin{tabular}{r|r|r|c|c|c|c|c|c}
\hline \hline
\multicolumn{2}{c|}{bin limits}        &  \multicolumn{1}{|c|}
{$A^{\gamma N\rightarrow D^0 \rm{X}}$}  &    
$\langle y\rangle$  &  $\langle Q^2\rangle$  & $\langle p_{\rm T}^D
\rangle$  &   $\langle E_{D}\rangle $  &$D(\langle X\rangle)$&
$a_{\rm LL}(\langle X\rangle) $\\
 $p_{\rm T}^{D}$(\gevc) &  $E_{D}$(GeV)         &         &           &   $(\GeV/c)^2$& (\gevc)       &   (GeV)         &     \\
\hline \hline
 0-0.3    & 0-30    &$ -1.34 \pm 0.85$ &  0.47 &  0.50 & 0.19 & 24.8&  0.57&  0.37 \\ 
 0-0.3    & 30-50   &$ -0.27 \pm 0.52$ &  0.58 &  0.75 & 0.20 & 39.2&  0.70&  0.48 \\ 
 0-0.3    & $>50$   &$ -0.07 \pm 0.66$ &  0.67 &  1.06 & 0.20 & 60.0&  0.80&  0.61 \\ 
 0.3-0.7  & 0-30    &$ -0.85 \pm 0.51$ &  0.47 &  0.47 & 0.50 & 25.1&  0.56&  0.26 \\ 
 0.3-0.7  & 30-50   &$ 0.09 \pm 0.29$ &  0.58 &  0.65 & 0.51 & 39.4&  0.71&  0.34 \\ 
 0.3-0.7  & $>50$   &$ -0.20 \pm 0.37$ &  0.67 &  0.68 & 0.50 & 59.6&  0.80&  0.46 \\ 
 0.7-1    & 0-30    &$ -0.47 \pm 0.56$ &  0.48 &  0.53 & 0.85 & 25.2&  0.58&  0.13 \\ 
 0.7-1    & 30-50   &$ -0.49 \pm 0.32$ &  0.58 &  0.66 & 0.85 & 39.1&  0.70&  0.17 \\ 
 0.7-1    & $>50$   &$ 1.23 \pm 0.43$ &  0.68 &  0.73 & 0.84 & 59.4&  0.81&  0.26 \\ 
 1-1.5    & 0-30    &$ -0.87 \pm 0.48$ &  0.50 &  0.49 & 1.21 & 25.7&  0.60&  0.01 \\ 
 1-1.5    & 30-50   &$ -0.24 \pm 0.25$ &  0.60 &  0.62 & 1.22 & 39.5&  0.73&  0.00 \\ 
 1-1.5    & $>50$   &$ -0.18 \pm 0.34$ &  0.69 &  0.77 & 1.22 & 59.3&  0.83&  0.04 \\ 
 $>1.5$   & 0-30    &$ 0.83 \pm 0.71$ &  0.52 &  0.51 & 1.77 & 26.2&  0.63&  $-0.13$ \\ 
 $>1.5$   & 30-50   &$ 0.18 \pm 0.28$ &  0.61 &  0.68 & 1.87 & 40.0&  0.74&  $-0.20$ \\ 
 $>1.5$   & $>50$   &$ 0.44 \pm 0.33$ &  0.71 &  0.86 & 1.94 & 59.9&  0.84&  $-0.24$ \\ 

\hline \hline
\end{tabular}
\caption{The asymmetries $A^{\gamma N\rightarrow D^0 \rm{X}}$ in  bins
  of $p_{\rm T}^{D^0}$ and $E_{D^0}$  
for the $D^{0}$ and $D^*$ sample combined, together with the 
averages of several kinematic variables. Only the statistical errors are given.
The relative systematic
uncertainty is 20\% which is 100\% correlated between the bins.}
\label{asym_results_d0ds}
\end{sidewaystable}

\section{Conclusion}

We have studied D$^0$ meson production in 160~GeV 
polarised muon scattering 
off a polarised deuteron target. The  $D^0$
decays into pairs of charged K and $\pi$ mesons were
selected using analysing the invariant mass
distributions of identified K$\pi$ pairs. Only one $D^0$ meson was demanded
in each event.

The data provide an average value of the gluon polarisation in
the nucleon, \DG, under the assumption that photon--gluon fusion
to a $c\bar c$ pair is the underlying partonic process
for open charm production, which is equivalent to a LO QCD approach.
 The result is 
\DG$=-0.49\pm 0.27(\mbox{stat})\pm 0.11(\mbox{syst})$
  at an 
average gluon momentum fraction, $\langle x\rangle \approx 0.11$
and at a scale $\mu^2\approx 13~(\GeV/c)^2$. This result is compatible with 
our previous result from the analysis of high-$p_{\rm T}$ hadron pairs 
but it is much less model dependent.  

The present measurement of the gluon polarisation in the nucleon, 
together with other measurements of COMPASS and HERMES, all
situated around $x\sim 0.1$, point towards a small gluon 
polarisation at that value of $x$. This is a hint for a small value
of the first moment, $\Delta G$, of the gluon helicity
distribution, although this in principle does not exclude a large value.

The longitudinal cross-section asymmetries
$A^{\gamma {N}\rightarrow {D}^0{\rm X}}$ were also extracted from our data 
and are presented in bins of the transverse momentum and the laboratory 
energy of the D$^0$. They may be used to constrain the values of 
$\Delta g(x)$ in future global NLO QCD analyses.

\section*{Acknowledgements}
We gratefully acknowledge the support of the CERN management and staff, the
special
effort of CEA/Saclay for the target magnet project, as well as
the skills and efforts of the technicians of the collaborating
institutes. 


%


\begin{thebibliography}{99.}
\raggedright
%
\bibitem{vernon} V.W. Hughes, Nucl.\ Phys.\ A {\bf 518} (1990) 371 and
references therein.
\bibitem{emc} EMC, J. Ashman {\it et al.}, Nucl.\ Phys.\ B {\bf 328} (1989) 1;
Phys.\ Lett.\ B {\bf 206} (1988) 364.
\bibitem{leader} E.~Leader, Spin in Particle Physics, Cambridge
  University Press (2001).
\bibitem{smc}SMC, B.~Adeva {\it et al.}, Phys.\ Rev.\ D
{\bf58} (1998) 112001.
\bibitem{compass} COMPASS, V.Yu. Alexakhin
{\it et al.}, Phys.\ Lett.\ B {\bf 647} (2007) 8. 
\bibitem{e155_d} E155, P.L.~Anthony {\it et al.},
Phys.\ Lett.\ B {\bf463} (1999) 339; see also list of references in
\cite{compass}.
\bibitem{hermes} HERMES, A.~Airapetian {\it et al.}, Phys.\ Rev.\ D {\bf
75} (2007) 012007; {\it erratum ibid.} D{\bf 76} (2007) 039901. 
\bibitem{jlab} CLAS, K.V.~Dharmawardane {\it et al.}, 
Phys.\ Lett.\ B {\bf 641} (2006) 11.
\bibitem{phenix} PHENIX, A.~Adare {\it et al.}, Phys.\ Rev.\ D {\bf 76}
(2007) 051106(R).
\bibitem{star} STAR, B.I.~Abelev {\it et al.}, 
Phys.\ Rev.\ Lett.\ {\bf 97} (2006) 252001.
\bibitem{refa8} J.~Ellis and R.~Jaffe, Phys.\ Rev.\ 9 {\bf D} (1974)
1444; {\it ibid.} {\bf 10} (1974) 1669.
\bibitem{bass} S.~Bass, The Spin Structure of the Proton, World
  Scientific Publishing (2007).
\bibitem{hermes_highpt} HERMES, A.~Airapetian {\it et al.},
Phys.\ Rev.\ Lett.\ {\bf 84} (2000) 2584; a smaller preliminary value from another method has been reported in P.~Liebig,
AIP Conf.\ Proc.\ {\bf 915} (2007) 331 (arXiv:0707.3617).
\bibitem{SMC_highpt} SMC, B.~Adeva {\it et al.}, Phys.\ Rev.\ D {\bf 70} (2004) 012002.
\bibitem{compass_highpt_lowq} COMPASS, E.S.~Ageev {\it et al.} 
Phys.\ Lett.\ B {\bf 633} (2006) 25; recently a new value from another method
has been reported in M.~Stolarski, Proc. of the XVIth Int. Workshop on
Deep-Inelastic Scattering, Eds. R.~Devenish and J.~Ferrando, London
(2008) (arXiv:0809.1803).
\bibitem{pythia} T.~Sj\"ostrand {\it et al.}, JHEP {\bf 0605} (1006) 026.
\bibitem{lepto} G.~Ingelman {\it et al.}, Comp.\ Phys.\ Comm.\ {\bf 101}
  (1997) 108.
\bibitem{emccharm} EMC, J.J.~Aubert {\it et al.}, Nucl.\ Phys.\ B {\bf 213}
  (1983) 31.
\bibitem{reanal} B.W.~Harris {\it et al.}, Nucl.\ Phys.\ B {\bf 461} (1996)
  181.
\bibitem{cernpp} COMPASS, M.~Alekseev {\it et al.},
  CERN-PH-EP/2008-003 (arXiv:0802.3023).
\bibitem{spectrometer} COMPASS, P.~Abbon {\it et al.},
Nucl.\ Instrum.\ Meth.\ A {\bf 577} (2007) 455.
\bibitem{beampol} N.~Doble {\it et al.}, Nucl.\ Instrum.\ Meth.\ A {\bf 343}
  (1994) 351.
\bibitem{compass_A1} COMPASS, E.S.~Ageev
{\it et al.}, Phys.\ Lett.\ B {\bf 612} (2005) 154. 
\bibitem{compass_lowq} COMPASS, V.Yu.~Alexakhin
{\it et al.}, Phys.\ Lett.\ B {\bf 647} (2007) 330. 
\bibitem{pretz}
%
  J.~Pretz and J.M.~Le Goff,
  Nucl.\ Instrum.\ Meth.\  A {\bf 602} (2009) 594
  (arXiv:0811.1426).
%
\bibitem{pretz_hab} For details see: J.~Pretz, habilitation thesis,
University of Bonn, Mathematisch-Naturwissenschaftliche Fakult\"at,
April 2007.
\bibitem{aroma} G.~Ingelman {\it et al.},
Comput.\ Phys.\ Commun.\ {\bf 101} (1997) 135;\\ see
http://www.isv.uu.se/thep/aroma/ for recent updates.
\bibitem{neuralnet} R.~Sulej {\it et al.}, Meas.\ Sci.\ Technol.\ {\bf 18}
  (2007) 2486.
\bibitem{phd-robinet} F.~Robinet, PhD thesis (Saclay), 
Univ.\ Paris Diderot (Paris 7), UFR de Physique, September 2008.
\end{thebibliography}
\end{document}